\documentclass[namedreferences]{SolarPhysics}
\usepackage[optionalrh]{spr-sola-addons} % For Solar Physics
\usepackage{graphicx}                    % For eps figures, newer & more powerfull
\usepackage{color}                       % For color text: \color command
\usepackage{url}                         % For breaking URLs easily trough lines
\newcommand{\aap}{    {\it Astron. Astrophys. }}

\newcommand{\apj}{    {\it Astrophys. J. }}

\newcommand{\solphys}{{\it Solar Phys. }}

\begin{document}

\begin{article}

\begin{opening}

\title{Investigation of quasi-periodic variations in hard X-rays of solar flares}

%%%%%%%%%%%%%%%%%%%%%%%%%%%%%%%%%%%%%%%%%%%%%%%%%%%
%% Authors Names
%
\author{J.~\surname{Jakimiec}\sep
        M.~\surname{Tomczak}%$^{1}$\sep
%        I.~\surname{}$^{2}$
       }

%%%%%%%%%%%%%%%%%%%%%%%%%%%%%%%%%%%%%%%%%%%%%%%%%%%
%% Runningheads
%
\runningauthor{J.\,Jakimiec \& M.\,Tomczak} \runningtitle{QPO in
HXRs of solar flares}

%%%%%%%%%%%%%%%%%%%%%%%%%%%%%%%%%%%%%%%%%%%%%%%%%%%
%% Affilations
%
  \institute{Astronomical Institute, University of Wroc{\l }aw,
  ul. Kopernika 11, 51-622 Wroc{\l }aw, Poland,
                     email: \url{jjakim; tomczak@astro.uni.wroc.pl}\\
%             $^{2}$ Second affiliation
%                     email: \url{e.mail-c} \\
             }

%%%%%%%%%%%%%%%%%%%%%%%%%%%%%%%%%%%%%%%%%%%%%%%%%%%
%%% Abstract
\begin{abstract}
The aim of the present paper is to use quasi-periodic oscillations
in hard X-rays (HXRs) of solar flares as a diagnostic tool for
investigation of impulsive electron acceleration. We have selected a
number of flares which showed quasi-periodic oscillations in hard
X-rays and their loop-top sources could be easily recognized in HXR
images. We have considered MHD standing waves to explain the
observed HXR oscillations. We interpret these HXR oscillations as
being due to oscillations of magnetic traps within cusp-like
magnetic structures. This is confirmed by a good correlation between
periods of the oscillations and the sizes of the loop-top sources.
We argue that a model of oscillating magnetic traps is adequate to
explain the observations. During the compressions of a trap
particles are accelerated, but during its expansions plasma, coming
from chromospheric evaporation, fills the trap, which explains the
large number of electrons being accelerated during a sequence of
strong impulses. The advantage of our model of oscillating magnetic
traps is that it can explain both the impulses of electron
acceleration and quasi-periodicity of their distribution in time.
\end{abstract}

%%%%%%%%%%%%%%%%%%%%%%%%%%%%%%%%%%%%%%%%%%%%%%%%%%%
%% Keywords
%

\end{opening}
%-------------------------------------------------

%%%%%%%%%%%%%%%%%%%%%%%%%%%%%%%%%%%%%%%%%%%%%%%%%%%
%% Sections
%
\section{Introduction}\label{intr}

Extreme ultraviolet (EUV) images from the TRACE satellite allowed to
observe oscillations of coronal loops during solar flares (see
\opencite{asc04} and papers refereed therein). Typical periods of
the oscillations were 200--300\,s. The oscillations have been
explained as standing MHD oscillations of whole coronal loops, i.e.
with nodes located at the loop footpoints.

\inlinecite{nak06} observed quasi-periodic oscillations (QPOs) with
similar periods, $P = 2-4$\,min, in hard X-rays (HXRs) of a solar
flare and similar periods, $P \approx 5$\,min were observed in soft
and hard X-rays of a solar flare by \inlinecite{fou05}. In both
those papers the observed periodicities were also interpreted as
being due to the oscillations of whole coronal loops.

On the other hand, in many flares QPOs with shorter periods, $P
\simeq 10-60$\,s, were observed in HXR emission (see
\opencite{lip78}). In the present paper we would like to show how
the investigation of these short-period oscillations can be used to
investigate impulsive electron acceleration.

In Sect.\,\ref{obs} observational data and their analysis are
presented. Sect.\,\ref{sec:disc} contains the discussion of our
results and Sect.\,\ref{concl} contains our conclusions.

In this paper we mostly investigate quasi-periodic HXR oscillations
with periods $P \simeq 10$--60\,s, but in Sect.\,\ref{llimb} we
present analysis of three limb flares with periods $P > 120$\,s.

\section{Observations and their analysis}\label{obs}

\begin{figure}
\centerline{\includegraphics[width=1\textwidth]{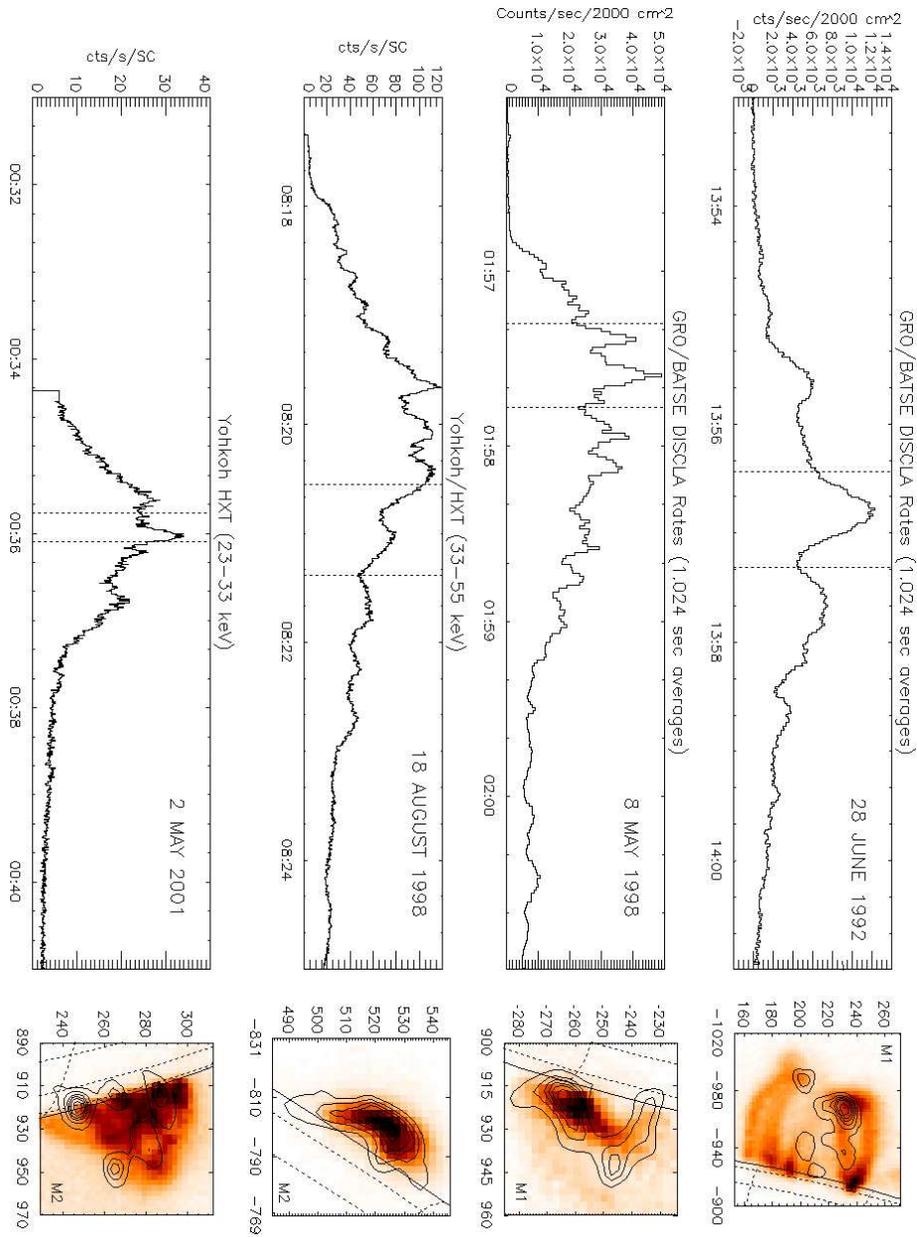}}
\caption{Hard X-ray observations for four limb flares. Left: HXR
light-curves (dashed vertical lines indicate accumulation times of
the HXR images shown at the right). Right: {\sl Yohkoh} HXR images
of the flares (isolines 0.12, 0.24, 0.35, 0.47, 0.59, and 0.78
$I_{max}$) and SXR images (gray scale). Solid line shows solar limb
and dashed lines are on the solar disc. The coordinates (x, y) are
in arcsec, the zero-point located at the disc center.} \label{limb1}
\end{figure}

We used HXR light-curves (recorded by {\sl Yohkoh} and {\sl Compton
Gamma Ray Observatory}) and HXR images (recorded by {\sl Yohkoh}).
First, from the {\sl Yohkoh} Hard X-ray Telescope Catalogue
\cite{sat06} we selected flares which showed at least three HXR
impulses and the time-intervals between the impulses were of similar
duration (see Figs.\,\ref{limb1} and \ref{limb2}). We have selected
about 50 appropriate flares. Next, we investigated their HXR images
and have taken for further analysis the flares for which the
altitude and diameter of the HXR loop-top source could be reliably
determined.

\subsection{Limb flares with periods $P = 10-60$\,s}
\label{slimb}

\begin{table}
\caption{List of limb flares with periods $P$ = 10--60\,s$^{\rm a}$}
\label{tab1}
\begin{tabular}{ccccccccc}
\hline
Flare & Date; time [UT] & $P$ & $h$ & $v_1$ & $B_1$ & $d$ & $v_2$ & $B_2$ \\
No. &  & [s] & [Mm] & [km\,s$^{-1}$] & [G] & [Mm] & [km\,s$^{-1}$] & [G] \\
\hline \hspace*{1.5mm}1$^{\rm b}$ & 92/06/28; 13:54--14:00 & 56; 66 & 36 & 4040 & 184 & \hspace*{-1.5mm}14.0 & 785 & 36 \\
2 & 98/05/08; 01:57--01:59 & 12; 18 & 18 & 9400 & 428 & 3.4 & 890 & 40 \\
3 & 98/08/18; 08:18--08:23 & 21; 28 & 10 & 3000 & 136 & 8.5 & \hspace*{-1mm}1270 & 57 \\
4 & 01/05/02; 00:35--00:37 & 24; (50) & 24 & 6300 & 290 & 5.1 & 670 & 31 \\
 \hline
\end{tabular}
\begin{list}{}{}
\item[$^{\rm a}$] Definition of parameters: $P$ are the time
intervals between HXR peaks; $h$ is the altitude of a HXR loop-top
source; $v_1$ is the velocity estimated according to Eq.\,(1); $B_1$
is the magnetic field strength calculated from Eq.\,(3); $d$ is the
diameter of a HXR loop-top source; $v_2$ is the velocity estimated
according to Eq.\,(4); $B_2$ is the magnetic field strength
calculated from Eq.\,(5). Time interval is the interval in which the
periods $P$ were determined.
\item[$^{\rm b}$] Structure and evolution of this flare was
investigated by \inlinecite{tom97}.
\end{list}
\end{table}

\begin{figure}
\centerline{\includegraphics[width=0.6\textwidth]{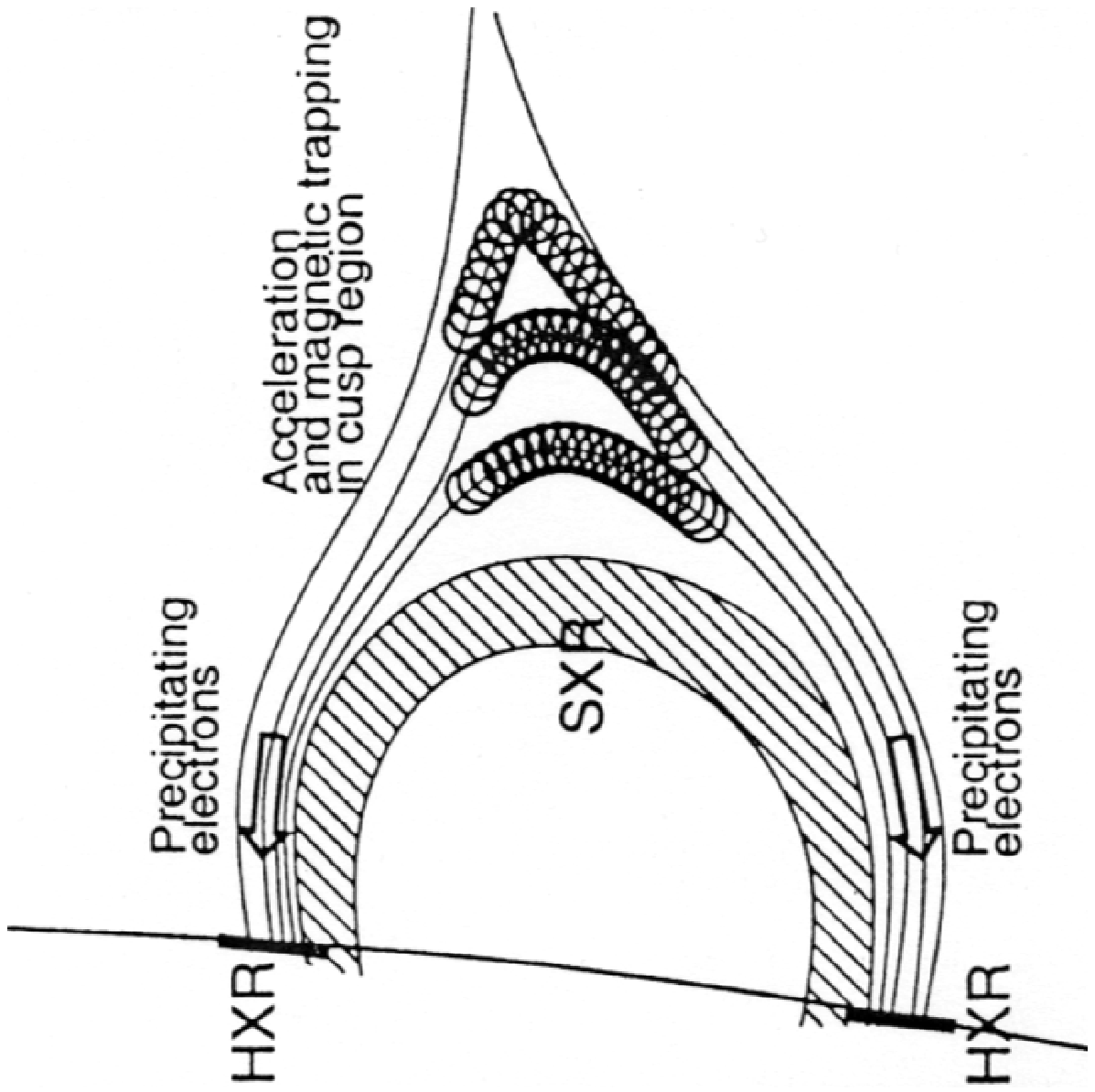}}
\caption{A cusp-like (triangular) magnetic structure according to
Aschwanden (2004). The tight spiral lines show motion of trapped
electrons.} \label{cusp}
\end{figure}

\begin{figure}
\centerline{\includegraphics[width=1\textwidth]{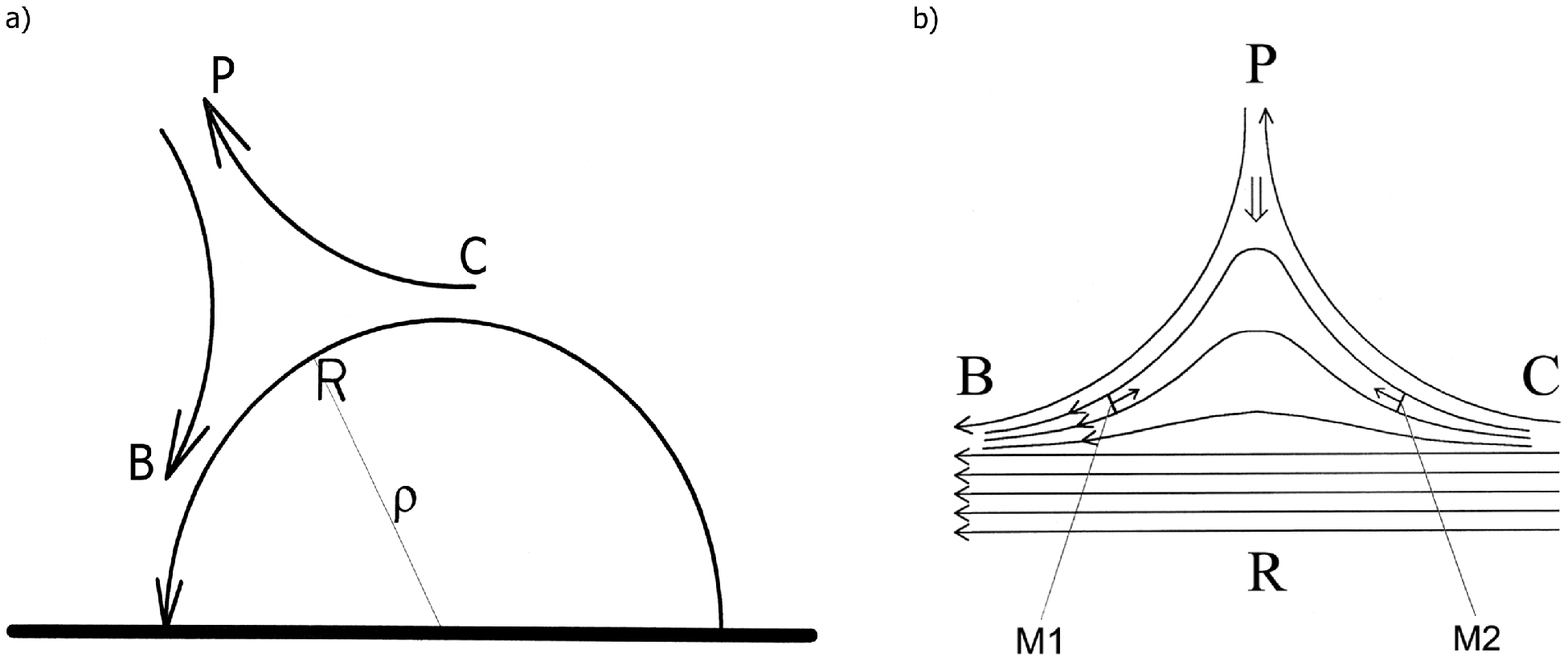}}
\caption{{\bf a} Schematic diagram showing a cusp-like magnetic
configuration. Magnetic fields PB and CP reconnect at P. The thick
horizontal line is the chromosphere. {\bf b} Detailed picture of the
BPC cusp-like magnetic structure.} \label{scheme}
\end{figure}

In this Section we investigate four limb flares. Their HXR
light-curves and images are shown in Fig.\,\ref{limb1}. We have
measured the loop-top source diameter, $d$, on various images of the
same HXR source and this way we have estimated its measuring (r. m.
s.) error to be about ${\pm}1.5$\,Mm. The loop-top source diameters
and their altitudes, $h$, are given in Table\,\ref{tab1}. In the
Table two values of period $P$ are given for each flare. They were
measured as the time-intervals between the strong HXR peaks. The
shorter period, which corresponds to higher magnetic field
compression, is taken in the estimates in Section\,2 [Eqs. (1) and
(4)]. By multiple measuring the same time interval, $P$, we have
estimated the r. m. s. errors of the measurements to be about
$\pm$2\,s for sharp peaks and about $\pm$4\,s for the peaks which
are not sharp or fluctuating (see error bars in Fig.\,\ref{plot}).

We have considered a hypothesis that the observed HXR oscillations
are the result of fast kink-mode oscillations (see below about their
excitation) and we have considered two kinds of magnetic structure
where they occur:
\begin{enumerate}
\item a simple magnetic loop (no inflexion points on the magnetic lines),
\item a magnetic loop with a ``cusp-like'' structure at the top
[Aschwanden's \shortcite{asc04} model -- see Fig.\,\ref{cusp}].
\end{enumerate}
In the first case the observed HXR QPOs are due to oscillations of
whole flaring loops, i.e. the nodes of standing waves are situated
at the loop footpoints. Assuming semi-circular shape of a loop, we
can calculate the velocity of the wave:
\begin{equation}
v_1 = \frac{2{\pi}{h}}{P}
\end{equation}
The values of $P$, $h$ and $v_1$ are given in Table\,\ref{tab1}.

For the MHD fast-mode oscillations it should be
\begin{equation}
v_1 \approx v_A = B_1/\sqrt{4{\pi}{\rho}},
\end{equation}
where $v_A$ is the Alfven speed and $B_1$ and $\rho$ are the mean
magnetic field and mass density inside the oscillating loop. Hence
\begin{equation}
B_1 \approx v_1 \sqrt{4{\pi}{\rho}}.
\end{equation}

\inlinecite{k+l08} estimated the electron number density, $N$, for a
number of HXR loop-top sources. Their values were concentrated
between 5$\times$10$^9$\,cm$^{-3}$ and 5$\times$10$^{10}$\,cm$^{-3}$
with a peak around 1.0$\times$10$^{10}$\,cm$^{-3}$. Using this
values of the electron density we have calculated values of the
magnetic filed strength $B_1$ which are given in Table\,\ref{tab1}.

\inlinecite{mar06} investigated oscillations of flaring loops using
Doppler-shifts in SXR spectra. The obtained distribution of periods
showed maximum between 3 and 5 minutes. He interpreted those
oscillations as being slow-mode or fast magnetosonic oscillations of
whole flaring loops.

Our periods, $P = 10-60$\,s, are significantly shorter than those
found by Mariska which suggests that the oscillations seen in HXRs
occur in smaller magnetic structures (see also comment in
Section\,\ref{sec:disc}). On the other hand, the cusp-like magnetic
structure is considered to be a ``standard'' magnetic structure in
solar flares (see \opencite{asc04}). Therefore we have assumed that
the oscillations seen in HXRs occur in such cusp-like structures.

Figure\,\ref{scheme}a shows a simple scheme of the cusp-like
structure of Fig.\,\ref{cusp}. Magnetic field lines steeply converge
around B and C. Charged particles are efficiently reflected there
(magnetic mirrors). Dynamics of the magnetic field inside the cusp
volume BPC is explained in Fig.\,\ref{scheme}b. Magnetic field PB
and CP reconnect at P. This generates a sequence of magnetic traps
which move downwards and collides with a stronger magnetic field R
below. The collision causes compression, the magnetic and gas
pressure increases inside the traps, so that the compression is
stopped, the traps can expand and perform magnetosonic oscillations.

The excited fast magnetosonic waves propagate along the magnetic
field lines, they meet regions of steep increase of Alfven speed
near B and C, where they are efficiently reflected back into the
trap. Hence, the magnetosonic waves will be efficiently trapped
within the magnetic traps and we can describe them as standing fast
magnetosonic (kink) oscillations trapped between B and C.

We can calculate the velocity of the waves
\begin{equation}
v_2 = 2 s/P, \end{equation} where $s$ is the distance between B and
C taken along the magnetic field lines. Diameters, $d$, of the HXR
loop-top sources have been used to estimate values of $s$: $2 s
\approx {\pi} d$. Hence
\begin{equation}
v_2 = \frac{{\pi}{d}}{P}
\end{equation}

Next we have calculated expected values of the magnetic field,
\begin{equation}
B_2 \approx v_2 \sqrt{4{\pi}{\rho}}.
\end{equation}
The values of $v_2$ and $B_2$ are given in Table\,\ref{tab1}. The
values show surprisingly low dispersion (see also values in
Table\,\ref{tab2}). We consider this to be a confirmation that the
oscillation are concentrated in cusp-like structures.

This model well describes generation of the short-duration HXR
impulses, see \citeauthor{jak02} (\citeyear{jak02},
\citeyear{jak05}). During the collision of the downwards moving
magnetic field with the magnetic field below, the magnetic traps
seen in Fig.\,\ref{scheme}b undergo compression and the magnetic
mirrors (M$_1$, M$_2$ in the Figure) move towards the center of the
trap. Particles contained in the traps are accelerated by betatron
effect and Fermi acceleration due to collisions with the converging
magnetic mirrors. Let us denote the trap ratio ${\chi} =
B_{max}/B_{min}$, where $B_{max}$ is the magnetic field strength at
the mirrors B and C and $B_{min}$ is the strength at the middle of a
trap. At the beginning of compression values of $\chi$ are high,
they decrease during the compression and accelerated particles can
escape towards the loop footpoints where they generate the HXR
footpoint emission. Some electrons emit HXRs before they escape from
the trap and this is why we can see the cusp-like volume BPC as a
HXR loop-top source. The parameter $\chi$ reaches its minimum value,
${\chi}_{min}$, during the maximum of compression. The values
${\chi}_{min}$ can be different for different traps seen in
Fig.\,\ref{scheme}b and the traps with low values of ${\chi}_{min}$
will be most efficient in particle acceleration and precipitation.
Traps which have higher values of ${\chi}_{min}$ are most efficient
in maintaining the oscillations.

Numerical simulations of this model of particle acceleration were
carried out by \inlinecite{k+k04}. A similar model of particle
acceleration was proposed by \inlinecite{s+k97}. The main difference
is that they assumed that the downward reconnection flow is
super-Alfvenic and therefore a fast shock is formed near the bottom
of the cusp-like volume. However, observations of \inlinecite{asa04}
show that the reconnection downflows can be sub-Alfvenic. Our model
of particle acceleration is valid also in the case of sub-Alfvenic
velocities in the cusp-like volume. \inlinecite{b+s05} have carried
out analytical calculations of particle acceleration in collapsing
magnetic traps.

\begin{table}
\caption{List of flares located on the solar disc$^{\rm a}$}
\label{tab2}
\begin{tabular}{ccccccccc}
\hline
Flare & Date; time [UT] & $P$ & $D$ & $v_1$ & $B_1$ & $d$ & $v_2$ & $B_2$ \\
No. & & [s] & [Mm] & [km\,s$^{-1}$] & [G] & [Mm] & [km\,s$^{-1}$] & [G] \\
\hline 5 & 91/10/31; 09:09--09:11 & 30; 39 & 21 & 2160 & 99 & 5.5 & 574 & 26 \\
\hspace*{1.5mm}6$^{\rm b}$ & 93/05/14; 22:04--22:05 & 25; 33 & 29 & 3600 & \hspace*{-1mm}166 & 5.3 & 660 & 30 \\
7 & 93/06/07; 14:20--14:22 & 60; 68 & 32 & 1650 & 76 & \hspace*{-1mm}13.7 & 717 & 33 \\
8 & 94/01/16; 23:16--23:18 & 51; 51 & 19 & 1180 & 54 & \hspace*{-1mm}12.3 & 757 & 35 \\
9 & 99/07/28; 08:13--08:14 & 31; 35 & 16 & 1660 & 76 & 6.8 & 690 & 32 \\
\hspace*{-1mm}10 & 00/06/23; 14:25--14:26 & 35; (76) & 17 & 1530 & 70 & 6.8 & 610 & 28 \\
\hspace*{-1mm}11 & 00/11/25; 18:38--18:39 & 21; 28 & 22 & 3270 & \hspace*{-1mm}150 & 4.1 & 613 & 28 \\
 \hline
\end{tabular}
\begin{list}{}{}
\item[$^{\rm a}$] Definition of parameters: $D$ is the distance
between the HXR footpoints; $v_1 = {\pi}D/P$. The other parameters
are the same as in Table\,\ref{tab1}.
\item[$^{\rm b}$] Structure and evolution of this flare was
investigated by \inlinecite{kol07}.
\end{list}
\end{table}

\begin{figure}
\centerline{\includegraphics[width=1\textwidth]{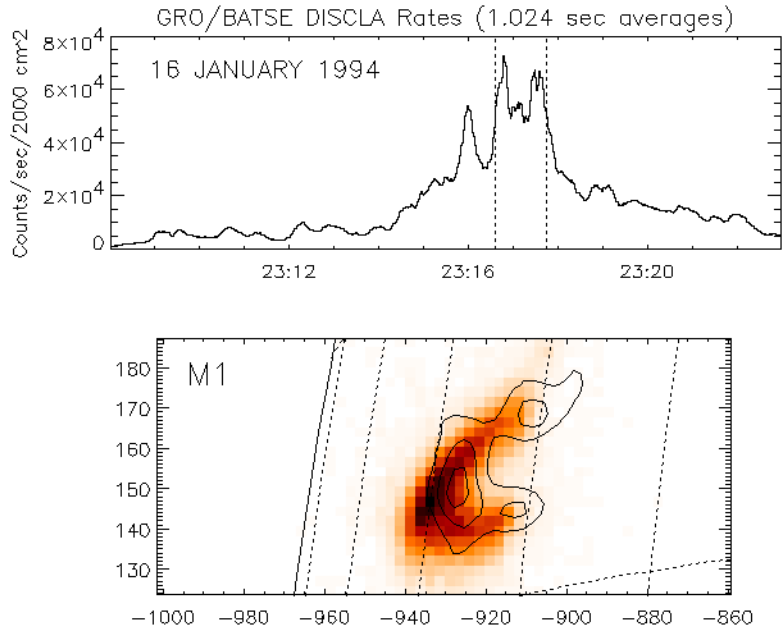}} \hfill
\caption{Observations of the flare of 16 January 1994. {\bf a}
GRO/BATSE light curve (nominally 25-50\,keV); {\bf b} {\sl Yohkoh}
HXT image, 23-33\,keV (isolines) and SXT image (gray scale).}
\label{img:disk}
\end{figure}

\subsection{Flares located on the solar disk}\label{sec:disk}

Usually, HXR footpoints sources are well seen in the case of flares
located on the solar disk. It is more difficult to recognize HXR
loop-top sources and to estimate their altitudes.

We have selected a sample of disk flares for which QPOs were
observed and loop-top sources could be identified and measured (see
example in Fig.\,\ref{img:disk}). The distance, $D$, between the HXR
footpoints has been taken for these flares as the measure of the
loop size ($h = D/2$ for a semicircular loop). We have calculated
wave velocity, $v_1$, for a whole loop as
\begin{equation}
v_1 = {\pi}D/P
\end{equation}
and the corresponding magnetic field, $B_1$, from Eq.\,(3).

Next, we have calculated the values of $v_2$ and $B_2$ for the
cusp-like model from Eqs.\,(4) and (5). The obtained values are
given in Table\,\ref{tab2}.

We see that Table\,\ref{tab2} confirms the properties seen in
Table\,\ref{tab1}. The values of $v_1$ and $B_1$ show large
dispersion. The values of $v_2$ and $B_2$ show much smaller
dispersion and they are in very good agreement with those in
Table\,\ref{tab1}. We consider this to indicate the advantage of the
cusp-like model.

\subsection{Investigation of sequences of HXR images}\label{sec:seq}

\begin{figure}
\centerline{\includegraphics[width=0.8\textwidth]{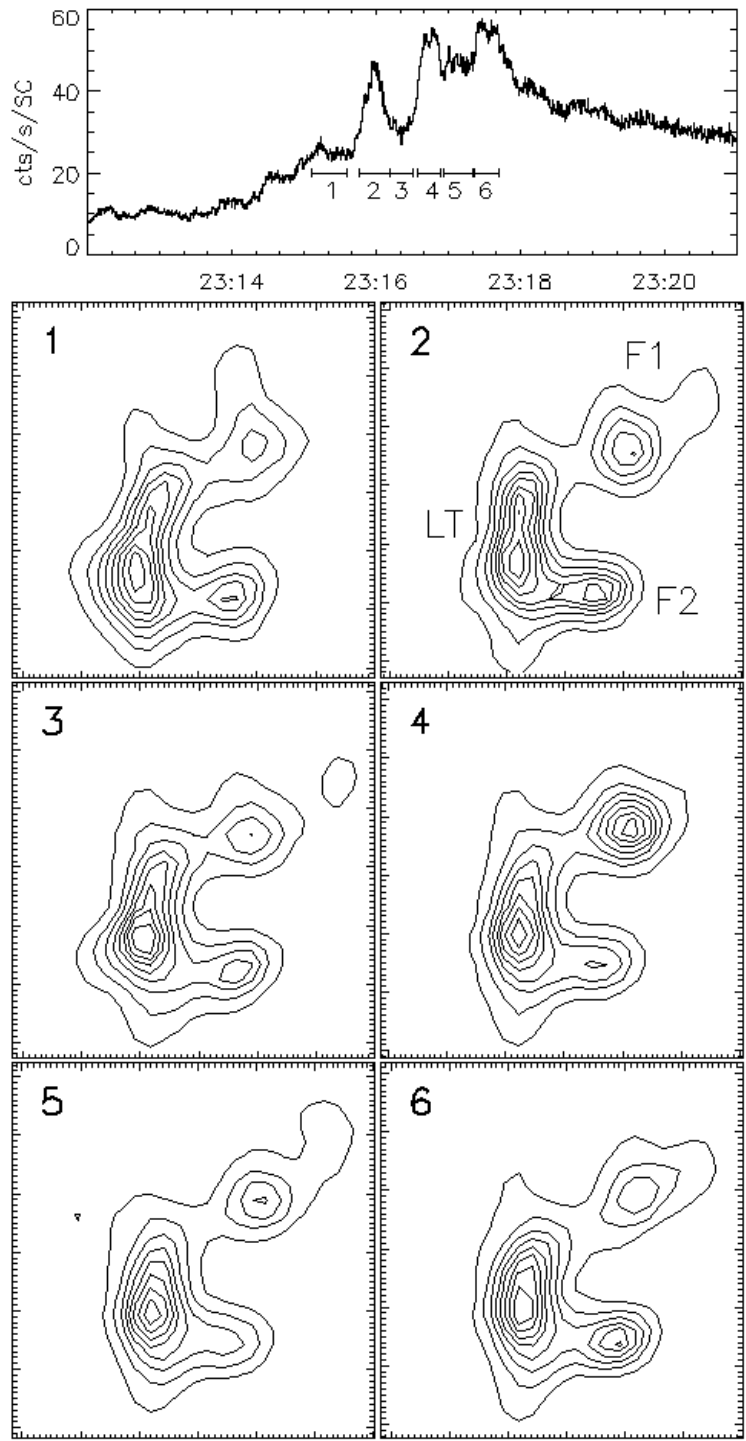}}
\caption{{\sl Yohkoh} 23-33 keV observations of the flare No\,8 (of
16 January 1994). Top: the HXR light-curve. The HXR images (lower
panels) have been accumulated for time intervals 1--6. See text for
discussion.} \label{img:seq}
\end{figure}

Important information can be derived from analysis of sequences of
HXR images. Such a sequence for a flare of 16 January 1994 is shown
in Fig.\,\ref{img:seq} ({\sl Yohkoh} 23-33\,keV images). The
location and structure of this flare are shown in
Fig.\,\ref{img:disk}. The loop-top (LT) and the footpoint (F1 and
F2) sources are well seen in the images of Fig.\,\ref{img:seq}. At
the top of this figure is shown the HXR light curve with the time
intervals for which the images Nos.\,1-6 have been accumulated.
Images Nos. 2, 4, and 6 correspond to the main HXR peaks and images
Nos. 3 and 5 correspond to ``valleys'' between the peaks.

Let us note that -- in spite of moderate spatial resolution -- a
trace of a triangular structure of the HXR loop-top source is seen
in images Nos.\,1 and 3 of the figure. It is not seen during the
maximal compression of the traps (images Nos.\,4 and 6).

\begin{figure}
\centerline{\includegraphics[width=1\textwidth]{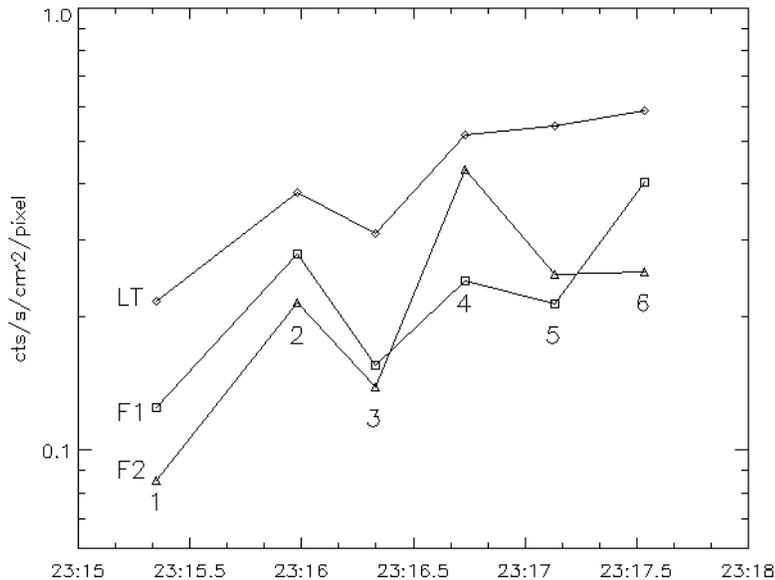}}
\hfill \caption{Time-variation of the HXR intensity (23--33\,keV) of
the brightest pixel in the footpoints F1 and F2 and in the loop-top
source.} \label{tvar}
\end{figure}

Figure\,\ref{tvar} shows time-variation of the intensity of the
brightest pixel in footpoints F1 and F2 and in loop-top source. We
see that the main HXR peaks, Nos.\,2, 4 and 6, are best seen in the
footpoint intensity, but they are less marked in the loop-top
intensity. This is in agreement with the outlined model: high
energies of electrons are achieved at the end of trap compression
and then they can easily escape towards the footpoints, so that
their contribution to the loop-top emission is moderate.

Let us note that there is some asymmetry in time-variation of the
footpoint intensity ($I_1 > I_2$ at peak No.\,4 and $I_1 < I_2$ at
peak No.\,6). This indicates that there is some asymmetry in the
oscillation of main trap (left-right oscillations in
Fig.\,\ref{scheme}b) which causes that the time variation of the
ratio $\chi$ is somewhat different for the two ends of the trap.

Similar results for other flares will be presented in another paper.

Hitherto we have considered impulses (peaks) of HXR emission. But it
is important that HXR emission from the loop-top and footpoint
sources is seen also during the HXR valleys (see
Figs.\,\ref{img:seq}-\ref{tvar}). This means that the process of
electron acceleration and precipitation is operating all the time
during the impulsive phase, also during the HXR valleys. To explain
this, let us return to Fig.\,\ref{scheme}b. Beginning of intensive
reconnection at P generates vigorous reconnection downflow which
collides with the transverse magnetic field R and causes compression
of a ``main'' magnetic trap and onset of its oscillations. But the
reconnection is continued and this means that a sequence of magnetic
traps, which are located above the main trap, undergo compression
and can oscillate. The compression and oscillations of these traps
are shifted in time relative to main trap, so that the acceleration
and precipitation of the electrons occurs during the whole impulsive
phase, including the HXR valleys. Hence, the HXR observations
suggest that the cusp-like volume BPC is a dynamic region which is
filled with moving, oscillating and colliding magnetic traps and it
is the volume where electrons are accelerated.

If the main oscillating trap is a dominant one (i.e. it contains
large number of particles), quasi-periodic sequence of HXR impulses
will be observed. If, however, there are several oscillating traps
of similar contribution to the HXR emission, the HXR peaks can be
randomly (i.e. not quasi-periodically) distributed, as it is the
case for impulsive phase of many flares.

\inlinecite{tom01} investigated positions of HXR and SXR loop-top
sources for limb flares, during their impulsive phase. He has found
that in most cases the HXR and SXR sources are co-spatial or nearly
cospatial. This result has been recently confirmed by
\inlinecite{k+l08} using {\sl RHESSI} observations; see also
\inlinecite{tom09}. These results indicate that in most cases both
the HXR and SXR loop-top emission, recorded during the impulsive
phase, comes from the same or nearly the same plasma volume. Let us
note that this common volume of loop-top HXR and SXR sources
confirms that HXR loop-top sources (like SXR ones) are also feeded
by chromospheric evaporation upflows.

\inlinecite{mas94} has found that in some flares the HXR loop-top
source is located clearly above the SXR loop -- see his flare of 13
January 1992 at 17:28 UT. The schematic diagram in Fig.\,\ref{cusp}
just concerns the ``Masuda flare'' of 13 January 1992.

In order to explain such cases as the Masuda flare we should take
into account that in the BPC cusp volume there is a sequence of many
moving traps and the dynamics of compression of individual traps is
determined by the time variation of the velocity, $v(t)$, of
reconnection outflow at the top of the cusp. The traps in the lower
part of the cusp volume are ``typical'': they undergo efficient
compression, i.e. ${\chi}_{min}$ is low for these traps. HXR
emission from these low traps was seen in the {\sl Yohkoh} L-channel
(14--23\,keV) -- see \inlinecite{mas94}. But these traps were not
seen at higher energies ($E > 20$\,keV) because (1) the higher
energies of electrons are achieved at the end of trap compression
and then they can immediately precipitate toward the loop
footpoints, (2) there was a dominant HXR ($E > 20$\,keV) sources at
higher altitude. The heating of the footpoints causes filling of the
loop with hot plasma and hence the loop is seen in SXRs. This
description is confirmed by the fact that HXR footpoint sources were
located just at the ends of the SXR loop (see Fig.\,12.12 in
\opencite{asc04}).

The enhanced HXR emission ($E > 20$\,keV) from the upper part of the
cusp indicates that the compression of the traps located there is
less efficient, so that the ratio $\chi$ remains high also during
maximum of compression and hence the escape of accelerated electrons
is more difficult (the traps are ``semiclosed''). The accelerated
electrons remain within the trap for longer time and they emit HXRs
before they escape from the trap.

Hence, the main point of our interpretation of the Masuda flare is
that the footpoint HXR sources are fed by the accelerated electrons
coming from ``typical'' traps, which are situated low in the cusp
volume BPC, and the top HXR source is a ``semiclosed'' magnetic
trap, from which escape of particles is more difficult.

Our model can also easily explain the cases when HXR loop-top source
is located {\sl below} the SXR loop. Then ``semiclosed'' traps are
located at the bottom of the BPC cusp volume, below the ``typical''
traps which are seen in SXRs. Such examples have been found by
\inlinecite{k+l08} and \inlinecite{tom09}. In their papers
${\Delta}h$ is the vertical distance between the centroids of HXR
and SXR loop-top sources and ${\Delta}h < 0$ means that the HXR
source is located below the SXR one (see Fig.\,2 in
\opencite{k+l08}). Let us note that such cases cannot be easily
explained by models which assume that a fast shock is formed within
the BPC cusp volume (\opencite{mas94}, \opencite{s+k97}).

\begin{figure}
\centerline{\includegraphics[width=1\textwidth]{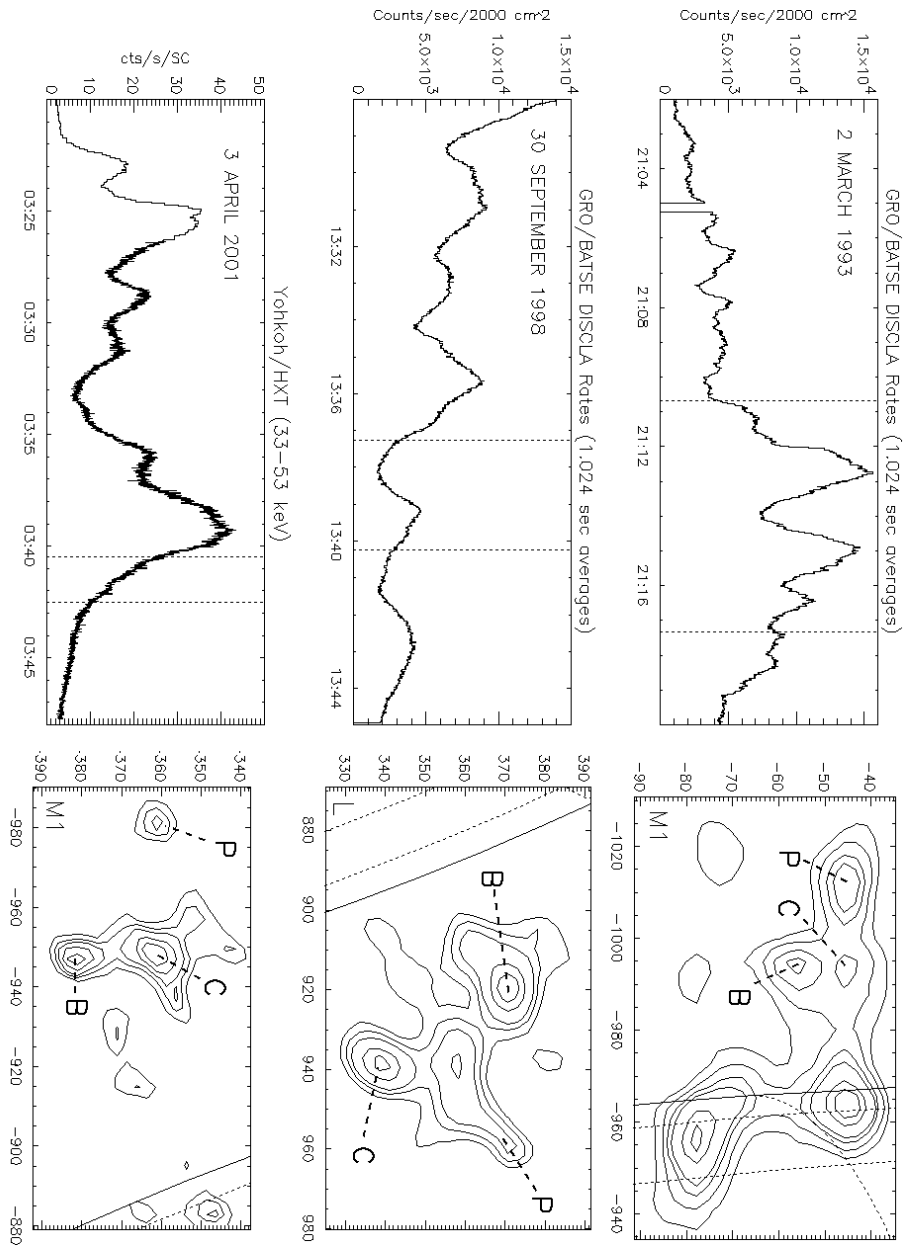}} \hfill
\caption{Hard X-ray observations for three limb flares. Left: HXR
light-curves (dashed vertical lines indicate accumulation times of
the HXR images shown on the right). Right: {\sl Yohkoh} HXR images
of the flares. Solid line shows solar limb and dashed lines are on
the solar disc.} \label{limb2}
\end{figure}

\subsection{Limb flares with periods $P > 120$\,s}
\label{llimb}

\begin{table}
\caption{Limb flares with periods $P > 120$\,s$^{\rm a}$}
\label{tab3}
\begin{tabular}{ccccccccc}
\hline
Flare & Date; time [UT] & $P$ & $h$ & $v_1$ & $B_1$ & $a$ & $v_2$ & $B_2$ \\
No. & & [s] & [Mm] & [km\,s$^{-1}$] & [G] & [Mm] & [km\,s$^{-1}$] & [G] \\
\hline \hspace*{-1.5mm}12 & 93/03/02; 21:10--21:15 & (93); 137 & 38 & 1700 & 78 & 15 & 690 & 32 \\
13$^{\rm b}$ & 98/09/30; 13:29--13:38 & 140; 166 & 38 & 1700 & 78 & 20 & 900 & 41 \\
14$^{\rm b}$ & 01/04/03; 08:18--08:23 & 150; 206 & 34 & 1400 & 64 & 20 & 840 & 39 \\
 \hline
\end{tabular}
\begin{list}{}{}
\item[$^{\rm a}$] Definition of parameters: $a$ is the size of cusp-like structure;
$v_2 = 2{\pi}a / P$; the other parameters are the same as in
Table\,\ref{tab1}.
\item[$^{\rm b}$] Structure and evolution of these flares were
investigated by \inlinecite{bak07}.
\end{list}
\end{table}

\begin{figure}
\centerline{\includegraphics[width=0.8\textwidth]{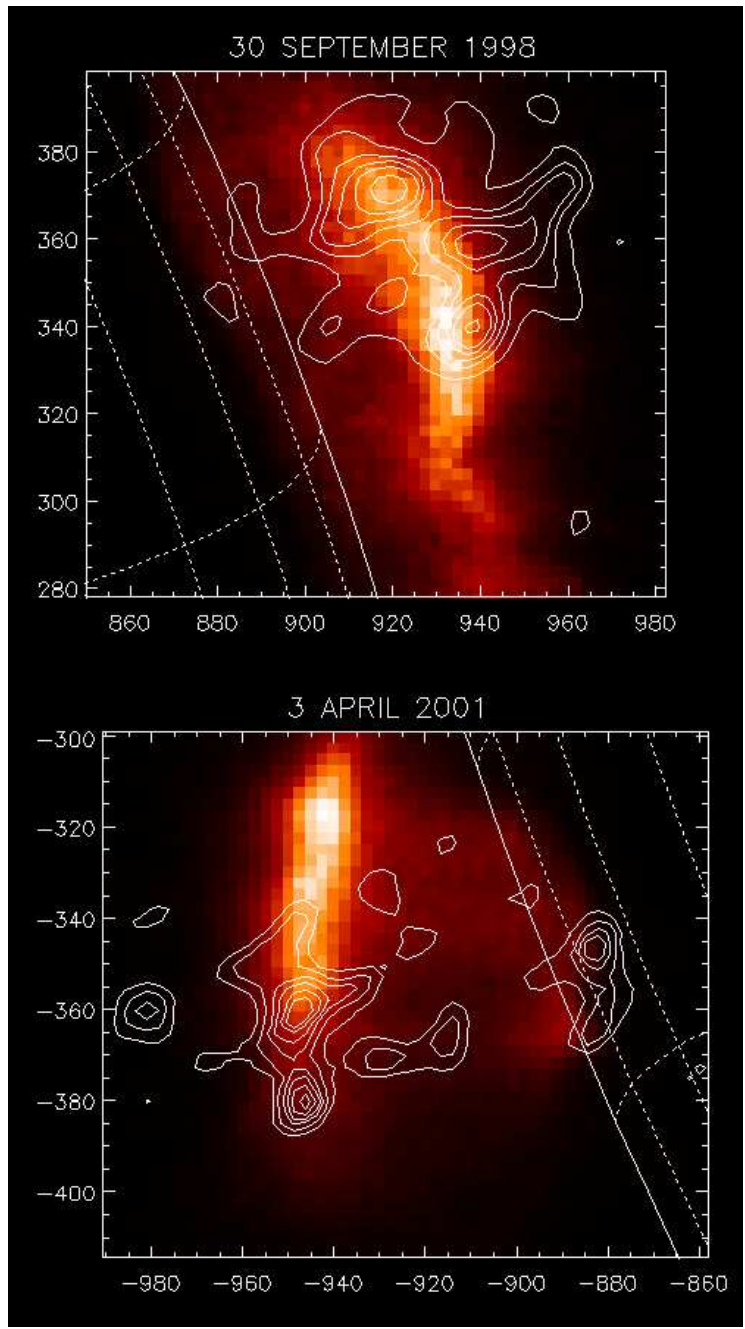}}
\hfill \caption{Soft X-ray images of the flares Nos.\,13 and 14
(gray scale). The isophotes show HXR images like in
Fig.\,\ref{limb2}: it is 14-23 keV image in the case of 30 September
1998 flare and 23-33 keV image for the 3 April 2001 flare.}
\label{limb:2ar}
\end{figure}

In this Section we consider three limb flares with periods $P >
120$\,s (Fig.\,\ref{limb2} and Table\,\ref{tab3}). All they are high
structures (see the altitudes, $h$, in Table\,\ref{tab3}) and
long-duration events (LDE). {\sl Yohkoh} SXR observations were
available for the flares Nos.\,13 and 14. Their SXR images (grey
scale) together with HXR isophotes are shown in
Fig.\,\ref{limb:2ar}. The SXR images show that they were big arcade
flares.

In particular, very interesting is the large triangular structure
seen in the HXR image of the flare No.\,13 of 30 September 1998 (see
Fig.\,\ref{limb2}). The dominant HXR sources, B and C, are located
at the altitude of SXR arcade (see Fig.\,\ref{limb:2ar}). Weak HXR
footpoint sources, located on the solar disc, are better seen in
{\sl Yohkoh} M1 (23-33 keV) and M2 (33-53 keV) images.

We have assumed that this triangular structure is analogous to BPC
cusp-like structure in Fig.\,\ref{scheme} where the observed
oscillations occur. The two dominant HXR sources, B and C,
correspond to the magnetic mirrors, B and C, in Fig.\,\ref{scheme}.
Charged particles spend somewhat longer time (calculated per unit
length along the magnetic trap) near the magnetic mirrors than in
other parts of the trap and therefore the HXR emission is enhanced
near the mirrors. The enhanced emission near P and along the line
running downwards from P, is due to the fact that plasma density is
somewhat higher there because of reconnection stream flowing from
the current sheet at P. The higher density increases the HXR
emissivity.

We have taken the distance, $a$, between the top source P and the
line connecting B and C, as the height of the cusp structure, and
calculated the velocity $v_2$ as
\begin{equation}
v_2 = \frac{2{\pi}a}{P}
\end{equation}
and then the magnetic field strength $B_2$ from Eq.\,(5). The values
of $v_1$ and $B_1$ calculated from Eqs.\,(1) and (3), as well as
$v_2$ and $B_2$ are given in Table\,\ref{tab3}.

Flare No.\,14 of 3 April 2001 is a similar case. It was also a large
arcade flare and the sources B and C were located at the altitude of
SXR arcade. A top HXR source P is seen in {\sl Yohkoh} M1
(23--33\,keV) images which confirms that this is also a similar
triangular HXR structure. HXR footpoint sources are seen on the
solar disc, but they are weaker than the B and C sources. Again, we
have measured the height of the cusp structure as the distance
between the top source P and the line connecting B and C and the
obtained values of $v_1$, $B_1$, $v_2$, and $B_2$ are given in
Table\,\ref{tab3}.

Our model (connection of sources B and C by a cusp-like structure
like in Fig.\,\ref{scheme}b) explains why in flares Nos.\,13 and 14
the sources B and C are similar in intensity and their time
evolution is similar. Figure\,\ref{limb:2ar} indicates that in the
both flares the plane BPC of the cusp-like structure is not
perpendicular to the arcade axis. This can be due to the shear in
footpoint motion at the photosphere (see Fig.\,3.10 in
\opencite{pri87}). The sources B and C are located at the places
where cusp-like magnetic lines meet the arcade loops (the places are
located on the opposite sides of the arcade axis).

No SXR images were available for the flare No.\,12 of 2 March 1993.
In the HXR image of this flare (Fig.\,\ref{limb2}) strong footpoint
sources are seen on the solar disc and a triangular structure BPC is
seen at the top. The strong footpoint emission indicates that the
oscillations of magnetic traps were more dynamic (low value of
${\chi}_{min}$) and therefore the accelerated electrons could easily
escape towards the footpoints. We have measured the altitude, $h$,
and the height, $a$, of the cusp-like structure at the top and
obtained values of $v_1$, $B_1$, $v_2$, and $B_2$ are given in
Table\,\ref{tab3}.

In Table\,\ref{tab3} we see that the values of $v_2$ and $B_2$ are
similar to those obtained for flares with periods 10--60\,s
(Tables\,\ref{tab1} and \ref{tab2}). This supports our assumption
that the large triangular HXR structures, considered in this
Section, are large magnetic cusp-like structures.

\section{Discussion}\label{sec:disc}

A lower limit for the Alfven speed can be estimated as follows:
\begin{equation}
v_A = 2.2 \times  10^{11} B/\sqrt{N} \quad\mbox{[cm\,s$^{-1}$]}\quad
\end{equation}
where $N$ is the electron density. For a magnetic trap it should be
\begin{equation}
B^2/(8 \pi) > 2 N k T,
\end{equation}
where $k$ is the Boltzmann constant and $T$ is the temperature. This
gives
\begin{equation}
B/\sqrt{N} > 8.3 \times 10^8 \sqrt{T}
\end{equation}
Hence the lower limit for the Alfven speed is [Eqs.~(9) and (11)]:
\begin{equation}
v_A > 1.8 \times 10^4 \sqrt{T} \quad\mbox{[cm\,s$^{-1}$]}\quad
\end{equation}
Taking $T > 10^7$~K for the temperature in the trap, we obtain
\begin{equation}
v_A > 5.7 \times 10^7 \quad\mbox{cm\,s$^{-1}.$}\quad
\end{equation}

We see in Tables\,\ref{tab1}--\ref{tab3} that our estimates of the
Alfven speed, $v_2$, are higher than this lower limit, but they are
not much higher. This means that the magnetic pressure, $B^2/(8
\pi)$, is not much higher than the gas pressure, $p$, in the
acceleration volume. This is important for MHD modeling of flaring
loops.

\begin{figure}
\centerline{\includegraphics[width=1\textwidth]{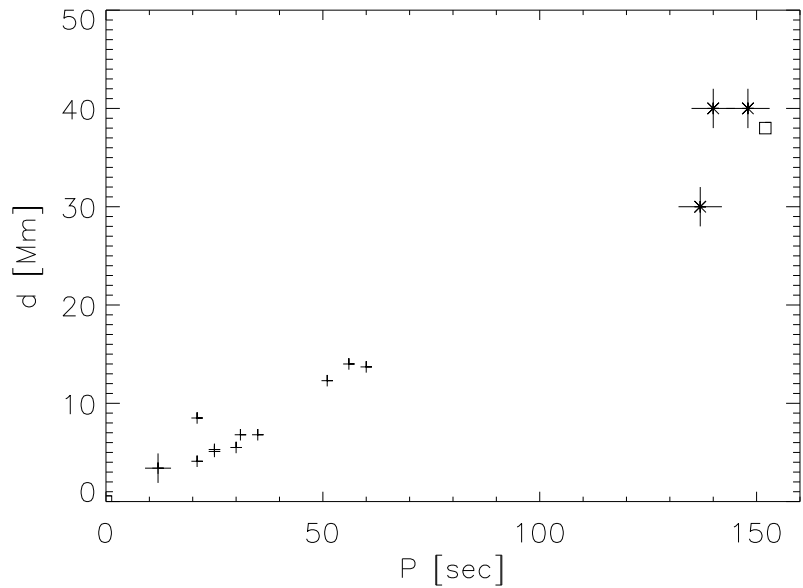}}
\hfill \caption{Correlation between the HXR loop-top source
diameter, $d$, and the oscillation period, $P$. For large sources
($P > 120$\,s) the values of $2a$ were taken for $d$ -- see text.
Point marked with a square relates to the flare investigated by
Ofman \& Sui (2006). The larger cross at the first data point on the
left shows r. m. s. error bars which concern all small HXR sources,
$P < 60$\,s, marked with small crosses.} \label{plot}
\end{figure}

In Fig.\,\ref{plot} the diameter, $d$, of the HXR loop-top source is
plotted vs. the oscillation period, $P$. For short periods, $P =
10-60$\,s, good correlation between $d$ and $P$ is seen. [A diagram
displaying $h$ (or $D$/2) vs. $P$ shows much larger dispersion.]
This is a confirmation that the investigated oscillations occur in
confined magnetic structures displayed as the HXR sources. This
supports our model of the oscillations of magnetic traps in
cusp-like magnetic structures.

For the large HXR sources ($P > 120$\,s), the values of $2a$ were
taken instead of $d$ in Fig.\,\ref{plot}, where $a$ is the height of
a triangular HXR structure. It appears that the large HXR structures
also fit the correlation between $d$ and $P$ determined by the small
structures. This supports our assumption that the large triangular
HXR structures are the cusp-like magnetic structures in which the
oscillations with period $P > 120$\,s occur.

\inlinecite{o+s06} have analyzed RHESSI observations of QPOs in a
flare of 19 January 2005 with period $P \simeq 150$\,s. We have
measured the size, $a$, of the HXR loop-top source in the same
manner as for our three large {\sl Yohkoh} flares and we have
plotted additional point marked with a square in our test diagram in
Fig.\,\ref{plot}. We see that this additional flare fits quite well
into our model of oscillating magnetic traps.

\begin{figure}
\centerline{\includegraphics[width=0.8\textwidth]{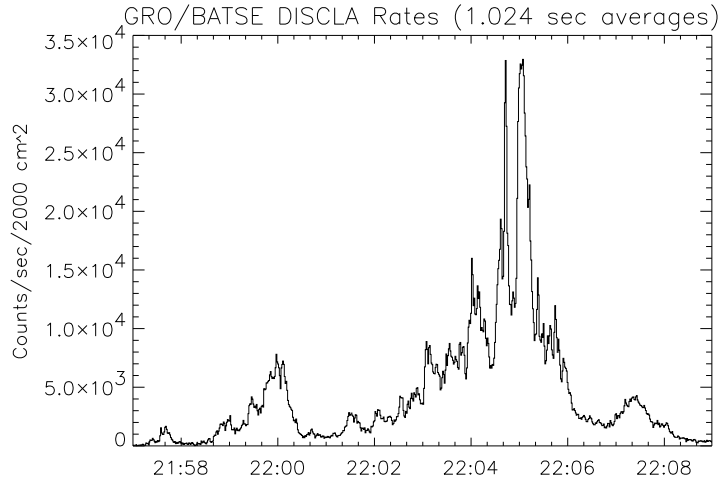}}
\hfill \caption{HXR light curve (nominally 25-50\,keV) for the flare
of 14 May 1993 (flare No.\,6).} \label{lc}
\end{figure}

It may be important that all four large flares in Fig.\,\ref{plot}
have periods about 140--150\,s, which suggests that they may be
overtones of the 300\,s solar oscillations. This would mean that the
oscillations are maintained due to a resonance with the 300\,s
oscillations. In particular, this seems to be probable in the case
of our flares Nos.\,13 and 14 and the flare investigated by
\inlinecite{o+s06}, in which the HXR oscillations are
quasi-sinusoidal (not typical sequences of increasing impulses like
in Fig.\,\ref{lc}).

For a number of flares \inlinecite{mar06} has found decaying
oscillations of periods 5.5$\pm$2.7 minutes in Doppler shifts of SXR
spectral lines. He explained them as being due to slow-mode or fast
magnetosonic oscillations of whole flaring loops. We have
investigated {\sl Yohkoh} HXR records for those flares and we have
found no counterpart of the Mariska's oscillations in the HXRs.
Mariska's flares were rather small, hence the observed oscillations
most probably were the relaxation oscillations of whole flare
magnetic structure which were excited by collision of violent
chromospheric evaporation upflows. Moreover, for some Mariska's
flares we have found HXR quasi-periodic oscillations of period
10--60\,s analogous to those investigated in the present paper. This
is an additional indication that the HXR oscillations with periods
10--60\,s occur in smaller volumes and they are different from the
longer-period oscillations observed by Mariska. These cases will be
investigated in a separate paper.

Important information can be obtained from {\sl CGRO}/BATSE
observations due to their high sensitivity in HXRs. An example of
such observations is shown in Fig.\,\ref{img:disk} (upper panel). We
see that a sequence of weak quasi-periodic impulses began about
23:09\,UT, but a typical impulsive phase began about 23:14\,UT.
(Similar weak oscillations before the impulsive phase are seen in
flare of 2 March 1993 in Fig.\,\ref{limb2}). These observations are
in agreement with the model of oscillating traps: The oscillations
began about 23:09\,UT. Accelerated particles hit the footpoints and
chromospheric evaporation begin.

The plasma moves along the same magnetic lines, connecting traps
with footpoints, and about 23:14\,UT it reaches the oscillating
traps. During the expansion of a trap this plasma fills the trap and
therefore a larger number of electrons will be accelerated during
the next compression of the trap. Hence, there is now a feedback
between the number of accelerated electrons and the density of the
plasma which reaches the traps. This feedback causes that the HXR
intensity increases with time, as it is seen in
Fig.\,\ref{img:disk}.

Hence, the model of oscillating magnetic traps provides a simple
explanation of the well-known paradox that the number of electrons,
which are accelerated during big flares, is much higher than it can
be provided by surrounding corona.

Let us repeat that, according to discussion of Fig.\,\ref{img:disk},
chromospheric evaporation begins a few minutes before the beginning
of the impulsive phase. The beginning of the evaporation causes
increase of the SXR emission. Hence, the model of electron
acceleration in oscillating magnetic traps is in agreement with the
well-known fact that SXR emission of a flare usually begins a few
minutes before the beginning of HXR impulsive phase.

Thermal conduction also contributes to the energy transport from the
loop top to its footpoints during the impulsive phase and it
increases the rate of chromospheric evaporation.

Our model of oscillating magnetic traps explains very well why the
time intervals, $P$, between the HXR impulses are not equal. The
oscillations occur in a volume of changing magnetic field and plasma
density, so that the Alfven speed can change from one pulse to
another.

In the present paper we have considered a cusp-like magnetic
configuration. We see from this example that two factors are
necessary for magnetic-trap compression and excitation of their
oscillations: a magnetic-field reconnection and an obstacle (a
transverse magnetic field) on the way of the reconnection flow. But
the location and orientation of the reconnection site can be
different in different flares.

\begin{figure}
\centerline{\includegraphics[width=0.8\textwidth]{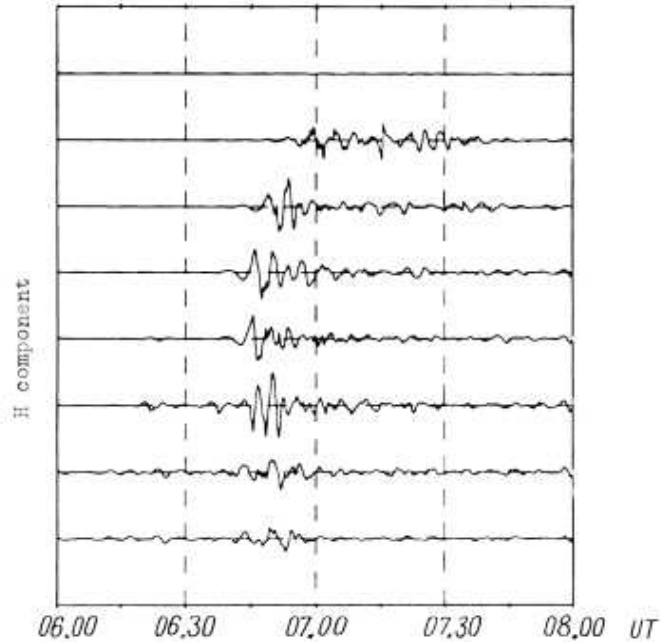}}
\hfill \caption{Magnetic field oscillations during the onset of a
geomagnetic substorm of 23 November 1970 (smooth variation has been
removed) as recorded by seven geomagnetic stations in Canada. The
distance between the horizontal lines amounts to 316 $\gamma$. From
Nishida (1978).} \label{sstorm}
\end{figure}

Let us note that our model of exciting oscillations of magnetic
traps is similar to that proposed for oscillations during onset of
magnetospheric substorms. The substorm begins with a violent
reconnection in the magnetospheric tail (see observations presented
by \opencite{nak04}). The reconnection outflow directed Earthward
collides with the inner (dipole-like) magnetic field, causes
magnetic field compression and excitation of oscillations. An
example of magnetic field oscillations during the onset of a
substorm is shown in Fig.\,\ref{sstorm} (a smooth substorm variation
of the magnetic field has been removed from these records). We see a
short sequence of oscillations which is similar to what we see in
HXRs of investigated flares (the smooth variation of HXR intensity
was not removed from our light curves).

\section{Summary and conclusions}\label{concl}

We have investigated flares for which quasi-periodic HXR
oscillations and HXR loop-top sources were clearly seen. Our
analysis has shown that the observed HXR oscillations are confined
within the loop-top sources. This is supported by a correlation
between the periods, $P$, and the sizes, $d$, of the loop-top
sources.

\inlinecite{mar06} has found that oscillations of whole flaring
loops had periods of 5.5$\pm$2.7 minutes. Recently we have found
that HXR oscillations with periods $P = 10-60$\,s occurred in some
of the Mariska's flares which confirms that the oscillations seen in
HXRs, occurred in smaller magnetic structures (these results will be
presented in a separate paper).

We have argued that a model of oscillating magnetic traps is
adequate to describe the HXR observations. During the compression of
a trap particles are accelerated, but during its expansion plasma
coming from chromospheric evaporation, fills the trap. This feedback
between the particle acceleration and filling of the traps explains
high number of electrons which are accelerated in strong flares.
This model is clearly supported by finding weak HXR oscillations a
few minutes before the impulsive phase (see Sect.\,\ref{sec:disc}).

We have estimated the mean strength of the magnetic field inside the
oscillating traps. It amounts to about 30--40 Gauss, which is lower
than assumed in other papers (see \opencite{asc04}).

We have compared the properties of the HXR oscillations with
available models of flare magnetic structure and particle
acceleration and indicated the most adequate scenario (see Sections
\ref{obs} and \ref{sec:disc}).

HXR light curves can be divided into two components (see
\opencite{asc96}, \opencite{tom01}): a smooth component (a smooth
line connecting valleys) and impulsive one (increases above the
smooth line). The smooth component is the result of continuous
acceleration in many traps whose oscillations are shifted in phase
(this is due to continuous reconnection at the top of cusp
structure). The increase of the smooth component during the
impulsive phase (like in Fig.\,\ref{lc}) is parallel to increase of
the SXR loop-top emission which confirms that the increase of the
smooth HXR component is also due to the increase of the density
within the cusp-like volume.

If there is a pronounced maximum in the time-profile of the
reconnection rate at the top of cusp, then a dominant (large) trap
will be generated within the cusp-like volume and its oscillations
will be seen as quasi-periodic oscillations in HXRs.

We have pointed out that in most cases the HXR and SXR sources are
cospatial or overlap during the impulsive phase which confirms that
both emissions come from the common cusp-like plasma volume. But in
seldom cases the HXR and SXR loop-top sources are clearly separated
(like in ``Masuda flare''). We explain such cases of HXR sources as
being due to magnetic traps, for which the ratio $\chi$ remains high
during the compression, so that accelerated electrons remain there
for a longer time and therefore generate enhanced HXR emission. This
explanation is also important for the flares having weak HXR
footpoint emission (like in the case of \opencite{v+b04}).

In our paper of 1998 \cite{jak98} we have found that (a) loop-top
flare kernels are multithermal, (b) significant random motions occur
in the kernels. On this basis we have proposed that the kernels are
filled with MHD turbulence. In the present paper we have used
observations of HXR oscillations to improve the model of loop-top
kernels. We have come to the conclusion that during the impulsive
phase the kernels are filled with oscillating magnetic traps. The
oscillations are excited and maintained by violent reconnection
flow. The advantages of this model are the following: (1) it
incorporates a simple model of particle acceleration (Fermi and
betatron acceleration), (2) the particles can easily escape after
their acceleration. We will observe this plasma volume as being
``turbulent'' (random motions due to oscillations shifted in phase;
temperature differences between different traps). (We think that MHD
turbulence can develop later on due to collision of chromospheric
evaporation upflows. This would ensure long duration of the
kernels.)

%% Figure
%
% \begin{figure}
% \centerline{\includegraphics[width=0.5\textwidth,clip=]{<fig.eps>}}
% \caption{}%\label{fig:?}
% \end{figure}

%% Table
%
% \begin{table}
% \caption{}%\label{tbl:?}
% \begin{tabular}{}
% \hline
% \multicolumn{2}{c}{<>}
% <data>
% \hline
% \end{tabular}
% \end{table}

%%%%%%%%%%%%%%%%%%%%%%%%%%%%%%%%%%%%%%%%%%%%%%%%%%%%%%%%%%%%%%%%%%%%%%%%%%%
%% Appendix
%
% \appendix
% \renewcommand{\theequation}{A\arabic{equation}}\setcounter{equation}{0}
% \renewcommand{\thefigure}{A\arabic{figure}}\setcounter{figure}{0}
% \renewcommand{\thetable}{A\arabic{table}} \setcounter{table}{0}

%%%%%%%%%%%%%%%%%%%%%%%%%%%%%%%%%%%%%%%%%%%%%%%%%%%%%%%%%%%%%%%%%%%%%%%%%%%
%% Acknowledgements
%
\begin{acks}
The {\sl Yohkoh} satellite is a project of the Institute of Space
and Astronautical Science of Japan. The {\sl Compton Gamma Ray
Observatory} is a project of NASA. The authors are thankful to Dr.
Urszula Bak-Ste$\acute{\rm{s}}$licka for her help in selecting
appropriate flares. This work was supported by Polish Ministry of
Science and High Education grant No. N\,N203\,1937\,33.
\end{acks}

%%% %%%%%%%%%%%%%%%%%%%%%%%%%%%%%%%%%%%%%%%%%%%%%%%%%%%%%%%%%%%
%% Bibliography
%
% Using BibTeX
%
% \bibliographystyle{spr-mp-sola}
% %\bibliographystyle{spr-mp-sola-cnd} %% Alternative style: no title, no concluding page
% \bibliography{<bib file>}

\begin{thebibliography}{}

   \bibitem[\protect\citeauthoryear{Asai {\it et al.}}{2004}]{asa04} Asai, A., Yokoyama, T., Shimojo, M., \&
   Shibata, K. 2004, \apj {\bf 605}, L77

   \bibitem[\protect\citeauthoryear{Aschwanden}{2004}]{asc04} Aschwanden, M.\,J. 2004, {\it Physics of the Solar
   Corona. An Introduction}, Springer: Praxis

   \bibitem[\protect\citeauthoryear{Aschwanden {\it et al.}}{1996}]{asc96} Aschwanden, M.\,J., Wills,
   M.\,J., Hudson, H.\,S., Kosugi, T., Schwartz, R.\,A. 1996, \apj
   {\bf 468}, 398

   \bibitem[\protect\citeauthoryear{Bak-Ste$\acute{\rm{s}}$licka}{2007}]{bak07}
   Bak-Ste$\acute{\rm{s}}$licka, U. 2007,
   {\it Ph.\,D. thesis}, University of Wroc{\l }aw

   \bibitem[\protect\citeauthoryear{Bogachev \& Somov}{2005}]{b+s05} Bogachev, S.\,A., \&
   Somov, B.\,V. 2005, {\it Astron. Letters} {\bf 31}, 537

   \bibitem[\protect\citeauthoryear{Foullon {\it et al.}}{2005}]{fou05} Foullon, C., Verwichte, E.,
   Nakariakov, V.\,M., \& Fletcher, L. 2005, \aap {\bf 440}, L59

   \bibitem[\protect\citeauthoryear{Jakimiec}{2002}]{jak02} Jakimiec, J. 2002, in Wilson E. (ed.), Proc.
   of the 10th European Solar Physics Meeting, {\it Solar Variability:
   From Core to Outer Frontiers},  ESA SP-{\bf 506}, 645

   \bibitem[\protect\citeauthoryear{Jakimiec}{2005}]{jak05} Jakimiec, J. 2005, in Danesy, D.,
   Poedts, S., De Groof, A., Andries, J. (eds.), Proc.
   of the 11th European Solar Physics Meeting, {\it The Dynamic Sun:
   Challenges for Theory and Observations},  ESA SP-{\bf 600}, 124.1

   \bibitem[\protect\citeauthoryear{Jakimiec {\it et al.}}{1998}]{jak98} Jakimiec, J., Tomczak, M., Falewicz, R.,
   Phillips, K.\,J.\,H., \& Fludra, A. 1998, \aap {\bf 334}, 1112

   \bibitem[\protect\citeauthoryear{Karlick$\acute{\rm{y}}$ \& Kosugi}{2004}]{k+k04} Karlick$\acute{\rm{y}}$, M.,
   \& Kosugi T. 2004, \aap {\bf 419}, 1159

   \bibitem[\protect\citeauthoryear{Ko{\l}oma$\acute{\rm{n}}$ski}{2007}]{kol07} Ko{\l}oma$\acute{\rm{n}}$ski, S.
   2007, \aap {\bf 465}, 1021

   \bibitem[\protect\citeauthoryear{Krucker \& Lin}{2008}]{k+l08} Krucker, S., \& Lin, R.\,P.
   2008, \apj {\bf 673}, 1181

   \bibitem[\protect\citeauthoryear{Lipa}{1978}]{lip78} Lipa, B. 1978, \solphys {\bf 57}, 191

   \bibitem[\protect\citeauthoryear{Mariska}{2006}]{mar06} Mariska, J.\,T. 2006, \apj {\bf 639},
   484

   \bibitem[\protect\citeauthoryear{Masuda}{1994}]{mas94} Masuda, S. 1994,
   {\it Ph.\,D. thesis}, University of Tokyo

   \bibitem[\protect\citeauthoryear{Nakamura}{2004}]{nak04} Nakamura, R. 2004, {\it ESA
   Bulletin} {\bf 118}, 64

   \bibitem[\protect\citeauthoryear{Nakariakov {\it et al.}}{2006}]{nak06} Nakariakov, V.\,M., Foullon, C.,
   Verwichte, E., \& Young, N.\,P. 2006, \aap {\bf 452}, 343

   \bibitem[\protect\citeauthoryear{Nishida}{1978}]{nis78} Nishida, A. 1978, {\it Geomagnetic
   Diagnosis of the Magnetosphere}, Physics and Chemistry in Space,
   New York: Springer

   \bibitem[\protect\citeauthoryear{Ofman \& Sui}{2006}]{o+s06} Ofman, L., \& Sui, L. 2006, \apj {\bf 644}, L149

   \bibitem[\protect\citeauthoryear{Priest}{1987}]{pri87} Priest, E.\,R.
   1987, {\it Solar Magnetohydrodynamics}, D.\,Reidel Publishing Company

   \bibitem[\protect\citeauthoryear{Sato {\it et al.}}{2006}]{sat06} Sato, J., Matsumoto, Y., Yoshimura, K.,
    et al. 2006, \solphys {\bf 236}, 351

   \bibitem[\protect\citeauthoryear{Somov \& Kosugi}{1997}]{s+k97} Somov, B. V., Kosugi, T. 1997, \apj {\bf 485},
   859

   \bibitem[\protect\citeauthoryear{Spitzer}{1962}]{spi62} Spitzer, L. 1962, {\it The Physics of
   Fully Ionized Gases}, Interscience, New York

   \bibitem[\protect\citeauthoryear{Tomczak}{1997}]{tom97} Tomczak, M. 1997, \aap {\bf 317}, 223

   \bibitem[\protect\citeauthoryear{Tomczak}{2001}]{tom01} Tomczak, M. 2001, \aap {\bf 366},
   294

   \bibitem[\protect\citeauthoryear{Tomczak}{2009}]{tom09} Tomczak, M. 2009,
   \aap {\bf 502}, 665

   \bibitem[\protect\citeauthoryear{Veronig \& Brown}{2004}]{v+b04} Veronig, A.\,M., \& Brown,
   J.\,C. 2004, \apj {\bf 603}, L117


\end{thebibliography}
%
% Without BibTeX

\end{article}
\end{document}